\documentclass{article}
\usepackage{spconf,amsmath,graphicx,hyperref}
\usepackage{makecell}
\usepackage{cite}
\usepackage{amsmath,amssymb,amsfonts}
\usepackage{float}
\usepackage{booktabs}
\usepackage{multirow}
\usepackage{enumitem}
\usepackage{algorithmic}
\usepackage{graphicx}
\usepackage{textcomp}
\usepackage{xcolor}
\usepackage{etoolbox}
\usepackage{subcaption}
\usepackage{threeparttable}

\title{TripleC Learning and Lightweight Speech Enhancement for Multi-Condition Target Speech Extraction}
\name{Ziling Huang$^{1,2 \dagger}$, Junnan Wu$^{2}$, Lichun Fan$^{2}$, Zhenbo Luo$^{2}$,  Jian Luan$^{2}$, Haixin Guan$^{3}$, Yanhua  Long$^{1,3}$\sthanks{Yanhua Long is the corresponding author. This work was sponsored by Natural Science Foundation of Shanghai (Grant No.25ZR1401277).\\$^{\dagger}$ Work done during internship at MiLM Plus, Xiaomi Inc.}}
\address{
  $^1$Shanghai Normal University, Shanghai, China\\
  $^2$MiLM Plus, Xiaomi Inc., Beijing, China\\
  $^3$Unisound AI Technology Co., Ltd. Beijing, China
  }

\begin{document}
\ninept

\maketitle

\begin{abstract}

In our recent work, we proposed Lightweight Speech Enhancement Guided Target Speech 
Extraction (LGTSE) and demonstrated its effectiveness in multi-speaker-plus-noise scenarios. 
However, real-world applications often involve more diverse and complex conditions, such as 
one-speaker-plus-noise or two-speaker-without-noise. 
To address this challenge, we extend LGTSE with a \textbf{C}ross-\textbf{C}ondition \textbf{C}onsistency 
learning strategy, termed \texttt{TripleC Learning}. This strategy is first validated under 
multi-speaker-plus-noise condition and then evaluated for its generalization across diverse scenarios. 
Moreover, building upon the lightweight front-end denoiser in LGTSE can flexibly process both noisy and 
clean mixtures and show strong generalization to unseen conditions, we integrate TripleC learning with a 
further proposed \texttt{parallel universal training} scheme that organizes batches containing multiple scenarios for the 
same target speaker. By enforcing consistent extraction across different 
conditions, easier cases can assist harder ones, 
thereby fully exploiting diverse training data and fostering a robust universal model. 
Experimental results on the Libri2Mix three-condition tasks demonstrate that the 
proposed LGTSE with TripleC learning achieves superior performance over 
condition-specific models, highlighting its strong potential for universal deployment in 
real-world speech applications. 
\end{abstract}


\vspace{-0.2cm}
\section{Introduction}
\label{sec:intro}

Target speech extraction (TSE) aims to extract the speech of a desired speaker from mixtures of interferers and background noise using an enrollment utterance, with applications in automatic speech recognition, hearing aids and speech communication. However, most existing approaches are designed and evaluated under a single condition, such as two-speaker mixtures or one-speaker-plus-noise, which limits their practicality. However, real-world applications often involve diverse and complex conditions, including one-speaker-plus-noise, two-speaker-without-noise and noisy multi-speaker scenarios. This discrepancy highlights the need for a universal TSE that can robustly generalize across multi-diverse conditions.

Recent studies on enrollment-guided TSE can be grouped into three categories: 
1) speaker embedding/encoder-based approaches 
\cite{pbsrnn,xsepformer,sdpccn,you2025investigation}, 
2) embedding/encoder-free methods \cite{sefnet,sefpnet,yangxue,hu2024smma,parnamaa2024personalized}, 
and 3) hybrid techniques that combine the two \cite{zhang2025multi}. 
However, research on universal TSE that generalizes across diverse conditions remains limited, 
especially for embedding-free models. For example, \cite{zeng2025useftse} proposed USEF-TSE, showing 
that embedding-free frameworks can integrate with different separation backbones. 
\cite{wang2023framework} introduced UPN, which unifies personalized and non-personalized TSE within a 
single model but relies on speaker embeddings. Similarly, \cite{huang2025unified} developed USEF-
PNet, enabling both personalized and non-personalized TSE, while being an embedding-free framework.
AnyEnhance \cite{anyenhance}, UniAudio \cite{uniaudio} and Metis \cite{metis} also offer unified 
frameworks supporting TSE and other speech tasks.
However, most of these studies only provide single-condition TSE performance, and they have not been 
evaluated under multiple interfering conditions, leaving their robustness unverified and 
highlighting the need for systematic studies on multi-condition TSE.

%

To the best of our knowledge, multi-condition TSE was first brought to significant 
attention in the community during the ICASSP 2021 DNS Challenge \cite{reddy2021icassp}, called 
personalized speech enhancement (PSE), and further explored in ICASSP 2022 \cite{dubey2022icassp} and 
2023 \cite{dubey2024icassp}. In these challenges, models were typically trained on data from all 
three conditions, but evaluation used mainly overall metrics, without analyzing each system across 
individual conditions. For example, top-ranking systems \cite{tea-pse,yan2023npu} were evaluated 
using overall performance without examining condition-specific performance. 
Later studies \cite{tea-pse2.0,ge2024dac} conducted condition-wise evaluations, 
showing that training on all conditions and adding innovations slightly improved overall performance, 
simpler scenarios were already well-handled, while the main limitation came from the challenging 
two-speaker-plus-noise condition, which still needs targeted improvement. 
Notably, most of these efforts focused on embedding/encoder-based models; in contrast, 
embedding/encoder-free TSE has seen few attempts to build a universal model handling multi-diverse 
conditions.

Building upon these observations, this work makes the following contributions: 
1) We extend our recently proposed LGTSE \cite{huang2025lightweight} by introducing a cross-condition consistency learning 
strategy, termed TripleC Learning, which is initially validated on multi-speaker-plus-noise 
mixtures and then generalized to multi-condition TSE tasks, 
showing consistent improvements in extraction performance;
2) We build a universal embedding-free TSE model for multiple conditions: a lightweight GTCRN-based \cite{rong2024gtcrn} 
denoiser in LGTSE flexibly handles both noisy and clean mixtures, and a parallel universal training strategy is proposed that 
organizes multiple conditions for the same target speaker within each batch, enabling effective integration with 
TripleC Learning; 
3) Experimental results on the Libri2Mix three-condition dataset demonstrate that the proposed universal model consistently 
outperforms condition-specific baselines, highlighting its robustness and practical potential for real-world deployment.

\begin{figure*}[t]
\centering
\includegraphics[width=0.9\textwidth]{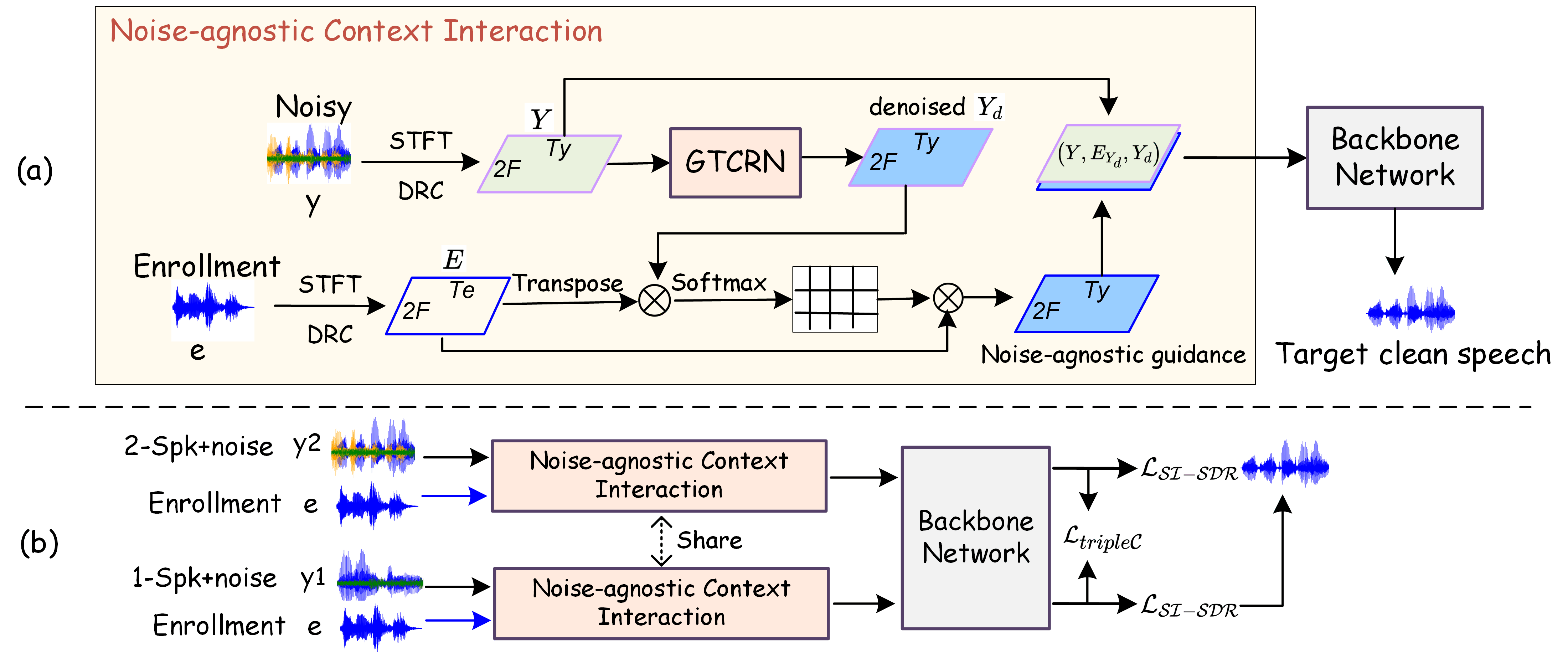}
\caption{Illustration of the proposed framework: (a) LGTSE architecture. (b) TripleC training scheme.}
\vspace{-0.3cm}
\label{fig:fig}
\end{figure*}

\section{Proposed Methods}
\label{sec:methods}
\subsection{Revisit LGTSE}

Fig.\ref{fig:fig}(a) shows our recently proposed LGTSE \cite{huang2025lightweight}, which is a lightweight speech enhancement-guided target speech extraction model. It consists of a front-end Noise-agnostic Context Interaction module and a back-end backbone. The model takes two inputs: noisy mixtures and enrollment speech, and outputs the extracted target clean speech. Both inputs are first transformed into complex time-frequency representations via short-time Fourier transform, with the enrollment \(\mathbf{E} \in \mathbb{R}^{2F \times T_e}\) and the noisy speech \(\mathbf{Y} \in \mathbb{R}^{2F \times T_y}\), where \(2F\) denotes the concatenated real and imaginary parts along the frequency axis, and \(T_e\) and \(T_y\) represent the number of frames of the enrollment and noisy speech, respectively.
Dynamic range compression \cite{drc} with compression factor \(\beta=0.5\) is applied to the magnitude spectrum:
\begin{equation}
\mathbf{E} = |\mathbf{E}|^\beta e^{j\theta_E}, \quad \mathbf{Y} = |\mathbf{Y}|^\beta e^{j\theta_Y}.
\vspace{-0.1cm}
\end{equation}

The noisy input \(\mathbf{Y}\) is then denoised by the GTCRN \cite{rong2024gtcrn} front-end to obtain \(\mathbf{Y}_d\), which is used for noise-agnostic context interaction:
\begin{equation}
\begin{aligned}
\mathbf{Y}_d &= \mathrm{GTCRN}(\mathbf{Y}), \\
\mathbf{E}_{Y_d} &= \mathbf{E} \times \mathrm{softmax}\left(\mathbf{E}^\mathrm{T} \times \mathbf{Y}_d\right).
\end{aligned}
\vspace{-0.1cm}
\end{equation}

The motivation is that directly using the noisy input would produce guidance contaminated by noise, which can degrade the target speech extraction performance.
In \cite{huang2025lightweight}, LGTSE showed competitive performance in multi-speaker noisy scenarios. Here, we further enhance the model and extend its generalization to diverse conditions.

\subsection{Cross-condition Consistency (TripleC) Learning}

Fig.~\ref{fig:fig}(b) illustrates the training scheme of our proposed Cross-condition Consistency (TripleC) Learning. 
To improve performance in multi-speaker-plus-noise conditions, we leverage the fact that single-speaker-plus-noise mixtures are easier for the model to extract, thus providing more reliable guidance. 
We therefore enforce consistency between outputs of the same target speaker with different conditional interfering and noise.

Specifically, let \(y_1\) denote a single-speaker-plus-noise mixture and \(y_2\) a two-speaker-plus-noise mixture, both paired with the same enrollment speech \(e\), and let \(s\) be the clean target speech. 
During training, the two input pairs \((y_1, e)\) and \((y_2, e)\) are fed in parallel. 
Each pair is processed by the noise-agnostic context interaction module and then passed into the backbone network. 
For brevity, we denote this end-to-end LGTSE mapping as \(f(\cdot)\), and the corresponding extracted signals are
\begin{equation}
\hat{s}_i = f(y_i, e), \quad i \in \{1,2\},
\end{equation}
where \(\hat{s}_1\) and \(\hat{s}_2\) are the estimated target speech signals from the two conditions, both aiming to approximate \(s\). Thus, the overall training objective combines consistency and supervision losses is defined as:
\begin{align}
\mathcal{L}_{\text{TripleC}} &= w \cdot \frac{1}{T} \sum_{t=1}^{T} \big| \hat{s}_1[t] - \hat{s}_2[t] \big|, \\
\mathcal{L}_{\text{SI-SDR}} &= - \sum_{i=1}^{2} \text{SI-SDR}(\hat{s}_i, s), \\
\mathcal{L}_{\text{total}} &= \mathcal{L}_{\text{SI-SDR}} + \mathcal{L}_{\text{TripleC}},
\end{align}
where \(T\) is the number of sampling points and \(w=50\).
Here, \(\mathcal{L}_{\text{TripleC}}\) enforces consistency between outputs under different conditions, while \(\mathcal{L}_{\text{SI-SDR}}\) ensures both estimates match the clean target \(s\). 
This joint optimization allows the model to exploit easier single-speaker mixtures to guide extraction in harder multi-speaker-plus-noise conditions, improving robustness and consistency.

\begin{figure*}[t]
\centering
\setlength{\abovecaptionskip}{0cm}
\includegraphics[width=\textwidth]{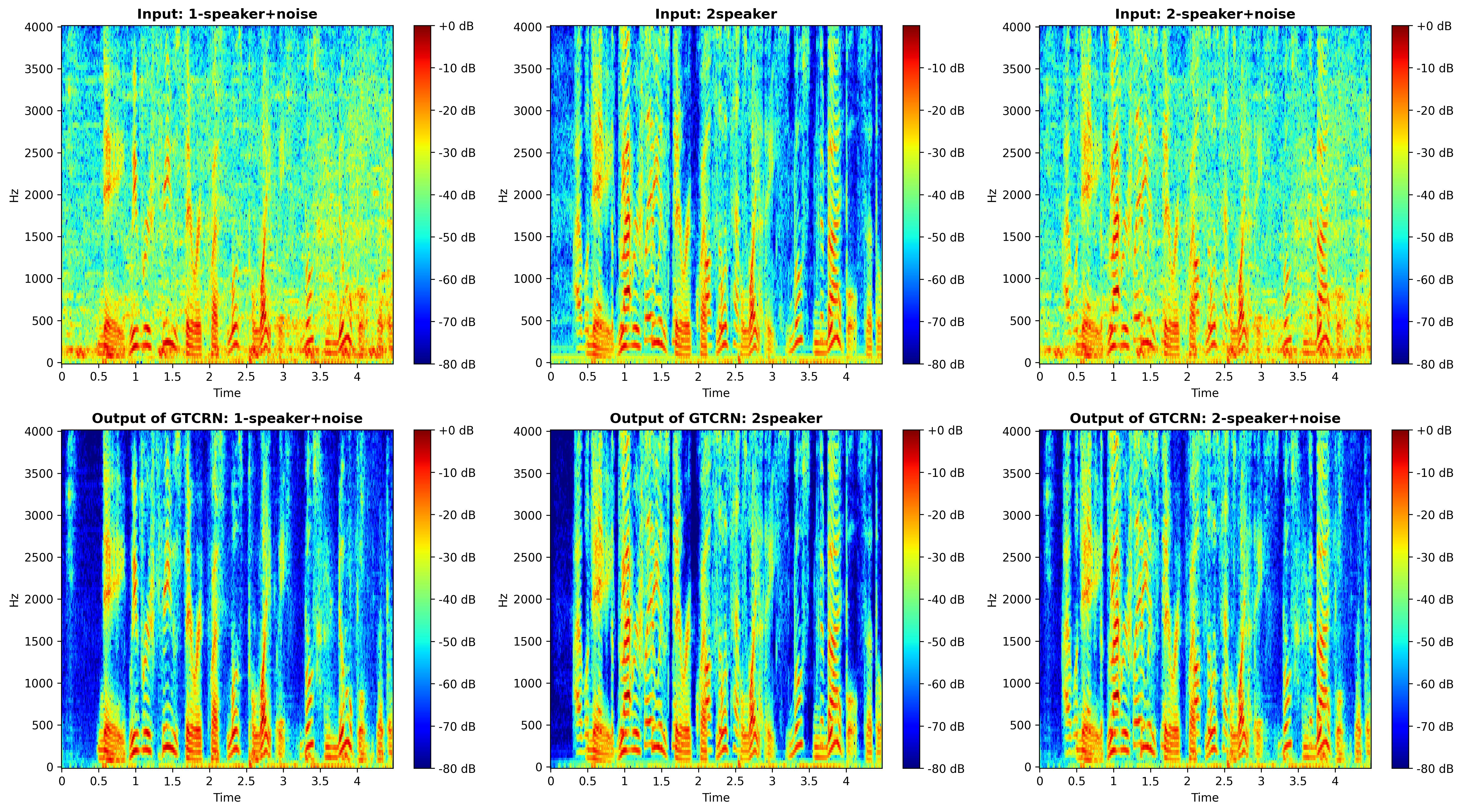}
\vspace{-0.2cm}
\caption{Illustration of GTCRN denoising performance. Top row: input mixtures under three conditions (single-speaker+noise, two-speaker clean, two-speaker+noise); bottom row: corresponding enhanced outputs.}
\label{fig:fig2}
\vspace{-0.4cm}
\setlength{\abovecaptionskip}{-0.5cm}
\end{figure*}

\subsection{TripleC-parallel Universal Training}
To extend TripleC Learning into a universal framework, we propose a parallel universal training scheme where each batch contains three types of mixtures sharing the same target speech and enrollment speech $e$: single-speaker-plus-noise, two-speaker-plus-noise, and two-speaker. 
These inputs are processed in parallel by the model, with each condition producing its own extracted output. 

The training objective combines two components.  
First, all outputs are supervised by their clean target speech using SI-SDR loss, ensuring accurate extraction across conditions.  
Second, the TripleC loss is applied only to the noisy conditions (single-speaker-plus-noise and two-speaker-plus-noise), aligning their outputs to let the easier case guide the harder one.  

This parallel setup allows the model to see all conditions simultaneously and to learn a shared representation that generalizes across them. 
The GTCRN front-end plays a crucial role: it naturally performs effective denoising when noise is present, while behaving almost as an identity mapping on clean mixtures (see Fig.~\ref{fig:fig2}). 
This property enables seamless universal training without any additional adjustment, laying a solid foundation for robust performance in diverse real-world scenarios.


\vspace{-0.1cm}
\section{Experiments and Results}
\label{ssec:exp}
\subsection{Datasets}

All our experiments are conducted on the publicly available Libri2Mix dataset \cite{libri2mix}, which covers three PSE conditions:  
1) 1-speaker+noise (\texttt{mix\_single}): a mixture containing only the target speaker and background noise;  
2) 2-speaker (\texttt{mix\_clean}): a mixture with a target speaker and one interfering speaker without additional noise;  
3) 2-speaker+noise (\texttt{mix\_both}): a mixture with a target speaker, one interfering speaker, and background noise.  
For clarity, we adopt the terms `1-speaker+noise', `2-speaker', and `2-speaker+noise' instead of the original \texttt{mix\_*} names in Libri2Mix. In each condition, the training set contains 13,900 utterances from 251 speakers, while the development and test sets each contain 3,000 utterances from 40 speakers.  
All mixtures are simulated in the `min' mode. Note that only the first speaker is taken as the target speaker, and all mixtures are resampled to 8~kHz, unless otherwise specified.

\begin{table*}[ht]
\renewcommand\arraystretch{1.4}
\centering
\caption{Performance comparison between our methods (E0–E4) and existing approaches (G0–G2). All metrics are the higher the better.}
\label{tab:tab_all}
\begin{threeparttable}
\begin{tabular}{p{0.5cm} p{2.0cm} p{2.6cm} ccc ccc ccc}
\toprule
ID & Training Data & Method 
   & \multicolumn{3}{c}{1-speaker+noise} 
   & \multicolumn{3}{c}{2-speaker} 
   & \multicolumn{3}{c}{2-speaker+noise} \\
\cmidrule(lr){4-6}\cmidrule(lr){7-9}\cmidrule(lr){10-12}
 & & & SI-SDR & PESQ & STOI & SI-SDR & PESQ & STOI & SI-SDR & PESQ & STOI \\
\midrule
-  & - & unprocessed  
   & 3.27 & 1.75 & 79.51 & -0.03 & 1.60 & 71.38 & -2.03 & 1.43 & 64.65 \\
\midrule
E0 & Condition-wise & SEF-PNet\cite{sefpnet}  
   & 14.50 & 3.05 & 92.47 & 13.00 & 3.05 & 89.71 & 7.43 & 2.14 & 80.31 \\
E1 & Condition-wise & LGTSE\cite{huang2025lightweight}  
   & \textbf{14.50} & \textbf{3.07} & \textbf{92.48} & 13.18 & 3.06 & 90.02 & 7.88 & 2.21 & 81.27 \\
\midrule
E2 & \makecell[l]{2-spk+noise \\ + 1-spk+noise} 
   & LGTSE+tripleC  
   & 14.25 & 3.03 & 92.03 & 11.87 & 2.81 & 88.44 & 8.41 & 2.32 & 82.41 \\
\midrule
E3 & \multirow{3}{*}{\makecell[l]{2-spk+noise \\ + 1-spk+noise \\ + 2-spk}} 
   & LGTSE+tripleC-parallel  
   & 14.28 & 3.03 & 92.15 & \textbf{13.33} & \textbf{3.07} & \textbf{90.19} & \textbf{8.58} & \textbf{2.33} & \textbf{82.61} \\
    \cmidrule(lr){3-12}
G0$^{\ddagger}$ & 
   & LGTSE+shuffled  
   & 14.42 & 2.28 & 92.91 & 15.01 & 2.57 & 92.53 & 9.70 & 1.72 & 85.52 \\
G1$^{\ddagger}$ & 
   & LGTSE+tripleC-parallel  
   & \textbf{14.51} & \textbf{2.30} & \textbf{93.10} & \textbf{15.23} & \textbf{2.64} & \textbf{92.95} & \textbf{10.12} & \textbf{1.77}& \textbf{85.96} \\
  \midrule
G2$^{\ddagger}$ & Condition-wise & DB-BSRNN\cite{wang2024enhancing} 
   & - & - & - & 13.84 & 2.53 & - & - & 1.77 & - \\
G3$^{*}$ & Condition-wise & NCSN++\cite{zhang2024ddtse} 
   & - & - & - & 13.80 & 2.24 & - & 9.70 & 1.55 & - \\
\bottomrule
\end{tabular}
\begin{tablenotes}
\footnotesize
\item[$^{\ddagger/ *}$] Results at 16~kHz on Libri2Mix-100 ($^{\ddagger}$) and Libri2Mix-360 (*), with two speakers alternating as the target speaker.
\end{tablenotes}
\end{threeparttable}
\vspace{-0.2cm}
\end{table*}

\vspace{-0.4cm}
\subsection{Models}

\textbf{GTCRN} \cite{rong2024gtcrn}, which has achieved state-of-the-art performance on  
speech enhancement tasks, has only 50K parameters and 0.03G MACs. \textbf{SEF-PNet} 
\cite{sefpnet}, used as one competitive embedding-free TSE baseline, has 6.08M parameters and 8.50G MACs. 
\textbf{LGTSE} \cite{huang2025lightweight}, which combines GTCRN and SEF-PNet, has 6.13M 
parameters and 8.53G MACs. As proposed in \cite{huang2025lightweight}, the LGTSE baseline 
is trained using a \texttt{pre-training+finetuning} two-stage training strategy: first, 
both the GTCRN front-end and backbone networks are pre-trained individually, 
then the whole LGTSE is fine-tuned to joint optimize the 
final system. In all the enhanced LGTSE+* experiments, the proposed TripleC and 
parallel universal training are all performed on both  the pre-training and 
finetuning stages. Full architecture details are provided in \cite{huang2025lightweight}.

\vspace{-0.3cm}
\subsection{Configurations}
We perform STFT using a 32 ms Hanning window with an 8 ms shift. 
The model is trained with Adam (initial LR = 0.0005), 
decayed by 0.98 every two epochs for the first 100 epochs and 
by 0.9 for the last 20 epochs. Gradient clipping is applied to limit 
the maximum L2-norm to 1. For evaluation, we report SI-SDR (dB)\cite{sisdr}, 
PESQ\cite{pesq}, and STOI (\%)\cite{stoi}.

\subsection{Results}
\subsubsection{Overall Results}

Table~\ref{tab:tab_all} presents the overall results of condition-wise and multi-condition scenarios (E0-E3). Condition-wise results refer to training and evaluating 
each condition separately. Compared with SEF-PNet (E0), LGTSE (E1) achieves 
similar results in the 1-speaker+noise case but shows clear gains under speaker interference. 
In the 2-speaker task, LGTSE improves SI-SDR and STOI, and in the more challenging 2-speaker+noise 
scenario, the gains are larger across all metrics. This confirms that the proposed noise-agnostic 
enrollment guidance in LGTSE provides more accurate cues in robust multi-condition TSE.

In E2, applying TripleC boosts the 2-speaker+noise performance over E1 significantly. 
The 1-speaker+noise condition shows a slight decrease, likely because the model 
sacrifices some performance on easier conditions to better align outputs for more 
challenging multi-speaker+noise mixtures, achieving overall improvement.
In contrast, the 2-speaker condition shows a noticeable performance drop, as the model has not seen clean speech without noise during training.

Extending TripleC to all conditions in E3 yields slight gains across scenarios. Parallel universal training encourages the model to process single-speaker+noise, multi-speaker clean, and multi-speaker+noise in the same feature space. This shared learning enhances the model’s ability to discriminate the target speaker, leading to minor improvements even for the multi-speaker clean condition, which is not directly constrained by TripleC.
Comparing E3 with E0 and E1, \textbf{\textit{an embedding-free single model achieves universal performance across conditions, and parallel  training with TripleC is generalizable}}. Except for a slight drop in 1-speaker+noise, all other conditions outperform models trained condition-wise independently, demonstrating the effectiveness of the unified training scheme.

\subsubsection{Comparison with Existing Approaches}

System G0–G3 in Table~\ref{tab:tab_all} presents the performance comparison with recent existing methods. For a fair comparison, all our models are trained on Libri2Mix-100 at 16kHz, with two speakers alternating as the target. G0 is taken as a standard baseline, where training data from the three conditions are randomly shuffled and fed into the model for training. G1 corresponds to our proposed model. It can be seen that applying the TripleC plus parallel universal learning strategy leads to consistent improvements across all conditions, demonstrating the effectiveness of our approach.  

Comparing G1 with other condition-wise trained state-of-the-art models (G2 and G3), LGTSE+TripleC-parallel achieves the highest performance. Specifically, under the 2-speaker condition, it achieves SI-SDR/PESQ/STOI = 15.23/2.64/92.95, and under the 2-speaker+noise condition, SI-SDR/PESQ/STOI = 10.12/1.77/85.96, surpassing the DB-BSRNN model. Moreover, our model even outperforms NCSN++, which is trained on Libri2Mix-360 with more than twice the amount of data.  
These results verify the effectiveness of a universal, multi-condition, embedding-free TSE model and highlight its potential for practical deployment in diverse acoustic scenarios.

\subsubsection{Effects of GTCRN in Multi-condition Tasks}

Fig.~\ref{fig:fig2} illustrates the denoising capability of our pre-trained LGTSE front-end, GTCRN. Although trained only on multi-speaker noisy mixtures, it generalizes well to unseen single-speaker noisy and multi-speaker clean conditions.
The first row shows the input mixtures: single-speaker+noise, two-speaker clean, and two-speaker+noise. The second row shows the corresponding GTCRN outputs. GTCRN effectively denoises single-speaker noisy inputs and preserves two-speaker clean inputs, removing any slight underlying noise. Performance on two-speaker noisy mixtures is also strong. These results demonstrate that GTCRN provides a robust foundation for a universal TSE model capable of handling both noisy and clean scenarios, enabling an embedding-free TSE system suitable for real-world multi-condition applications.

\section{Conclusion}

In this work, we built upon our previous LGTSE model and proposed TripleC Learning, a cross-condition consistency strategy to 
improve performance on multi-speaker-plus-noise mixtures. We first 
validated its effectiveness, and further, by adopting a parallel model  
training scheme, we developed a universal embedding-free TSE model 
capable of handling multiple conditions. The effectiveness of our 
extraction method was verified on the Libri2Mix three-condition 
datasets. Future work includes more generalization experiments, optimizing the model for low-latency, 
real-time deployment in practical applications.

\clearpage
\vfill\pagebreak
\bibliographystyle{IEEEbib}
\bibliography{strings,refs}

\begin{thebibliography}{10}

\bibitem{pbsrnn}
J.~Yu, H.~Chen, Y.~Luo, R.~Gu, and C.~Weng,
\newblock ``High fidelity speech enhancement with band-split {RNN},''
\newblock in {\em Proc. Interspeech}, 2023, pp. 2483--2487.

\bibitem{xsepformer}
K.~Liu, Z.~Du, X.~Wan, and H.~Zhou,
\newblock ``{X-SEPFORMER}: End-to-end speaker extraction network with explicit optimization on speaker confusion,''
\newblock in {\em Proc. ICASSP}, 2023, pp. 1--5.

\bibitem{sdpccn}
J.~Han, Y.~Long, et~al.,
\newblock ``{DPCCN}: Densely-connected pyramid complex convolutional network for robust speech separation and extraction,''
\newblock in {\em Proc. ICASSP}, 2022, pp. 7292--7296.

\bibitem{you2025investigation}
Z.~You, Z.~Zhou, L.~Li, and D.~Wang,
\newblock ``An investigation on speaker augmentation for end-to-end speaker extraction,''
\newblock {\em arXiv preprint arXiv:2505.21805}, 2025.

\bibitem{sefnet}
B.~Zeng, H.~Suo, Y.~Wan, and M.~Li,
\newblock ``{SEF-Net}: Speaker embedding free target speaker extraction network,''
\newblock in {\em Proc. Interspeech}, 2023, pp. 3452--3456.

\bibitem{sefpnet}
Z.~Huang, H.~Guan, H.~Wei, and Y.~Long,
\newblock ``{SEF-PNet}: Speaker encoder-free personalized speech enhancement with local and global contexts aggregation,''
\newblock in {\em Proc. ICASSP}, 2025, pp. 1--5.

\bibitem{yangxue}
X.~Yang, C.~Bao, J.~Zhou, and X.~Chen,
\newblock ``Target speaker extraction by directly exploiting contextual information in the time-frequency domain,''
\newblock in {\em Proc. Interspeech}, 2024, pp. 10476--10480.

\bibitem{hu2024smma}
Y.~Hu, H.~Xu, Z.~Guo, H.~Huang, and L.~He,
\newblock ``{SMMA-Net}: An audio clue-based target speaker extraction network with spectrogram matching and mutual attention,''
\newblock in {\em Proc. ICASSP}, 2024, pp. 1496--1500.

\bibitem{parnamaa2024personalized}
T.~P{\"a}rnamaa and A.~Saabas,
\newblock ``Personalized speech enhancement without a separate speaker embedding model,''
\newblock in {\em Proc. Interspeech}, 2024, pp. 4863--4867.

\bibitem{zhang2025multi}
K.~Zhang, J.~Li, S.~Wang, Y.~Wei, Y.~Wang, Y.~Wang, and H.~Li,
\newblock ``Multi-level speaker representation for target speaker extraction,''
\newblock in {\em Proc. ICASSP}, 2025, pp. 1--5.

\bibitem{zeng2025useftse}
B.~Zeng and M.~Li,
\newblock ``{USEF-TSE}: Universal speaker embedding free target speaker extraction,''
\newblock {\em IEEE Transactions on Audio, Speech, and Language Processing}, 2025.

\bibitem{wang2023framework}
Z.~Wang, R.~Giri, D.~Shah, J.-M. Valin, M.~M. Goodwin, and P.~Smaragdis,
\newblock ``A framework for unified real-time personalized and non-personalized speech enhancement,''
\newblock in {\em Proc. ICASSP}. IEEE, 2023, pp. 1--5.

\bibitem{huang2025unified}
Z.~Huang, H.~Guan, and Y.~Long,
\newblock ``Unified architecture and unsupervised speech disentanglement for speaker embedding-free enrollment in personalized speech enhancement,''
\newblock {\em arXiv preprint arXiv:2505.12288}, 2025.

\bibitem{anyenhance}
J.~Zhang, J.~Yang, Z.~Fang, Y.~Wang, Z.~Zhang, Z.~Wang, F.~Fan, and Z.~Wu,
\newblock ``{AnyEnhance}: A unified generative model with prompt-guidance and self-critic for voice enhancement,''
\newblock {\em arXiv preprint arXiv:2501.15417}, 2025.

\bibitem{uniaudio}
D.~Yang, J.~Tian, X.~Tan, R.~Huang, S.~Liu, H.~Guo, X.~Chang, J.~Shi, J.~Bian, Z.~Zhao, et~al.,
\newblock ``{UniAudio}: Towards universal audio generation with large language models,''
\newblock in {\em Proc.ICML}, 2024.

\bibitem{metis}
Y.~Wang, J.~Zheng, J.~Zhang, X.~Zhang, H.~Liao, and Z.~Wu,
\newblock ``{Metis}: A foundation speech generation model with masked generative pre-training,''
\newblock {\em arXiv preprint arXiv:2502.03128}, 2025.

\bibitem{reddy2021icassp}
C.~K.~A. Reddy, H.~Dubey, V.~Gopal, R.~Cutler, S.~Braun, H.~Gamper, R.~Aichner, and S.~Srinivasan,
\newblock ``{ICASSP 2021} deep noise suppression challenge,''
\newblock in {\em Proc. ICASSP}, 2021, pp. 6623--6627.

\bibitem{dubey2022icassp}
H.~Dubey et~al.,
\newblock ``{ICASSP 2022} deep noise suppression challenge,''
\newblock in {\em Proc. ICASSP}. IEEE, 2022, pp. 9271--9275.

\bibitem{dubey2024icassp}
H.~Dubey, A.~Aazami, V.~Gopal, B.~Naderi, S.~Braun, R.~Cutler, A.~Ju, M.~Zohourian, M.~Tang, M.~Golestaneh, et~al.,
\newblock ``{ICASSP 2023} deep noise suppression challenge,''
\newblock {\em IEEE Open Journal of Signal Processing}, vol. 5, pp. 725--737, 2024.

\bibitem{tea-pse}
Y.~Ju, W.~Rao, X.~Yan, Y.~Fu, et~al.,
\newblock ``{TEA-PSE}: Tencent-ethereal-audio-lab personalized speech enhancement system for {ICASSP} 2022 {DNS} challenge,''
\newblock in {\em Proc. ICASSP}, 2022, pp. 9291--9295.

\bibitem{yan2023npu}
X.~Yan, Y.~Yang, Z.~Guo, L.~Peng, and L.~Xie,
\newblock ``The {NPU-Elevoc} personalized speech enhancement system for {ICASSP 2023 DNS Challenge},''
\newblock in {\em Proc. ICASSP}, 2023, pp. 1--2.

\bibitem{tea-pse2.0}
Y.~Ju, S.~Zhang, W.~Rao, et~al.,
\newblock ``{TEA-PSE 2.0}: Sub-band network for real-time personalized speech enhancement,''
\newblock in {\em Proc. SLT}, 2023, pp. 472--479.

\bibitem{ge2024dac}
X.~Ge, J.~Han, H.~Guan, and Y.~Long,
\newblock ``Dynamic acoustic compensation and adaptive focal training for personalized speech enhancement,''
\newblock {\em Applied Acoustics}, vol. 216, pp. 109803, 2024.

\bibitem{huang2025lightweight}
Z.~Huang, J.~Wu, L.~Fan, Z.~Luo, J.~Luan, H.~Guan, and Y.~Long,
\newblock ``Lightweight speech enhancement guided target speech extraction in noisy multi-speaker scenarios,''
\newblock {\em arXiv preprint arXiv:2508.19583}, 2025.

\bibitem{rong2024gtcrn}
X.~Rong, T.~Sun, X.~Zhang, Y.~Hu, C.~Zhu, and J.~Lu,
\newblock ``{GTCRN}: A speech enhancement model requiring ultralow computational resources,''
\newblock in {\em Proc. ICASSP}, 2024, pp. 971--975.

\bibitem{drc}
A.~Li, C.~Zheng, R.~Peng, and X.~Li,
\newblock ``On the importance of power compression and phase estimation in monaural speech dereverberation,''
\newblock {\em JASA Express Letters}, p. 014802, 2021.

\bibitem{libri2mix}
J.~Cosentino, M.~Pariente, S.~Cornell, A.~Deleforge, and E.~Vincent,
\newblock ``{LibriMix}: An open-source dataset for generalizable speech separation,''
\newblock {\em arXiv preprint arXiv:2005.11262}, 2020.

\bibitem{wang2024enhancing}
J.~Wang, S.~Wang, J.~Li, K.~Zhang, Y.~Qian, and H.~Li,
\newblock ``Enhancing speaker extraction through rectifying target confusion,''
\newblock in {\em Proc. SLT}, 2024, pp. 349--356.

\bibitem{zhang2024ddtse}
L.~Zhang, Y.~Qian, L.~Yu, H.~Wang, H.~Yang, S.~Liu, L.~Zhou, and Y.~Qian,
\newblock ``{DDTSE}: Discriminative diffusion model for target speech extraction,''
\newblock in {\em Proc. SLT}, 2024, pp. 294--301.

\bibitem{sisdr}
J.~Le~Roux, S.~Wisdom, H.~Erdogan, and J.~R. Hershey,
\newblock ``{SDR}–half-baked or well done?,''
\newblock in {\em Proc. ICASSP}, 2019, pp. 626--630.

\bibitem{pesq}
M.~Wang, C.~Boeddeker, R.~Dantas, A.~Seelan, et~al.,
\newblock ``{PESQ} (perceptual evaluation of speech quality) wrapper for python users,'' Zenodo, 2022.

\bibitem{stoi}
C.~H. Taal, R.~C. Hendriks, J.~Heusdens, and R.~Jensen,
\newblock ``A short-time objective intelligibility measure for time-frequency weighted noisy speech,''
\newblock in {\em Proc. ICASSP}, 2010, pp. 4214--4217.

\end{thebibliography}

\end{document}